\documentclass[review]{elsarticle}

\usepackage{lineno,hyperref}
\usepackage{url}
\usepackage[utf8x]{inputenc}
\modulolinenumbers[5]

\journal{Journal of \LaTeX\ Templates}

\begin{document}

\begin{frontmatter}

\title{The Number of Confirmed Cases of Covid-19 by using Machine Learning: Methods and Challenges}

\author[in]{Amir Ahmad\corref{cor1}}
\ead{amirahmad@uaeu.ac.ae.com}
\address[in]{College of Information Technology, United Arab Emirates University, Al Ain, UAE}

\cortext[cor1]{Corresponding author}
\author[uk]{Sunita Garhwal}

\author[santosh]{Santosh Kumar Ray}

\author[gagan]{Gagan Kumar}
\author[kau]{Sharaf J. Malebary}
\author[kau]{Omar Mohammed Omar Barukab}

\address[uk]{Department of Computer Science and Engineering, Thapar University, Patiala, India.}
\address[santosh]{Department of Information Technology, Khawarizmi International College, Al Ain, UAE}
\address[gagan]{Department of Physics, Indian Institute of Technology Guwahati, Guwahati, 781039, Assam, India}
\address[kau]{Faculty of Computing and Information Technology, P.O. Box 411, King Abdulaziz University, Rabigh 21911, Jeddah, Saudi Arabia}

\begin{abstract}
Covid-19 is one of the biggest health challenges that the world has ever faced. Public health policy makers need the reliable prediction of the confirmed cases in future to plan medical facilities. Machine learning methods learn from the historical data and make a prediction about the event. Machine learning methods have been used to predict the number of confirmed cases of Covid-19. In this paper, we present a detailed review of these research papers. We present a taxonomy that groups them in four categories. We further present the challenges in this field. We provide suggestions to the machine learning practitioners to improve the performance of machine learning methods for the prediction of confirmed cases of Covid-19.
\end{abstract}


\end{frontmatter}

\section{Introduction}
Coronavirus disease 2019 (COVID-19) is an infectious disease caused by severe acute respiratory syndrome coronavirus 2 (SARS-CoV-2) \cite{Whocovid}. It was first reported in Wuhan, China in December 2020 \cite{Whochina}. Since then it has spread all over the world. It was declared pandemic by World Health Organization (WHO) on $11^{th}$ March 2020 \cite{Whopandemic}. As on $4^{th}$ June 2020, there were more than 6.5 million COVID-19 confirmed cases across 188 countries \cite{Johnhop}.  
    The fast spread of the Covid-19 has put a lot of pressure on healthcare systems of countries. Predicting the number of confirmed cases
in future has become an important task for the public health policy makers so that they can increase their medical facilities accordingly. It spreads differently in countries (Fig. 1 and Fig. 2). Different governments propose various measures such as lockdown, social distancing, closing of schools etc. to slow down the spread of Covid-19. Effect of different responses can be measured with accurate prediction models so that the response can be modified to be more effective.
\begin{figure}
	\centering
	\includegraphics[width=100mm]{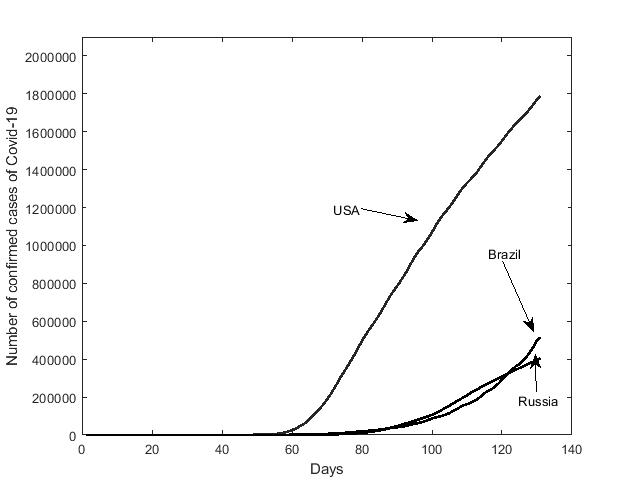}
	\caption{Confirmed cases of three countries (USA, Brazil and Russia) where the number of confirmed cases is increasing steadily. From $22^{nd}$ January 2020 to $31^{st}$ May 2020 \cite{Johnhopdata}}
\end{figure} 

\begin{figure}
	\includegraphics[width=170mm]{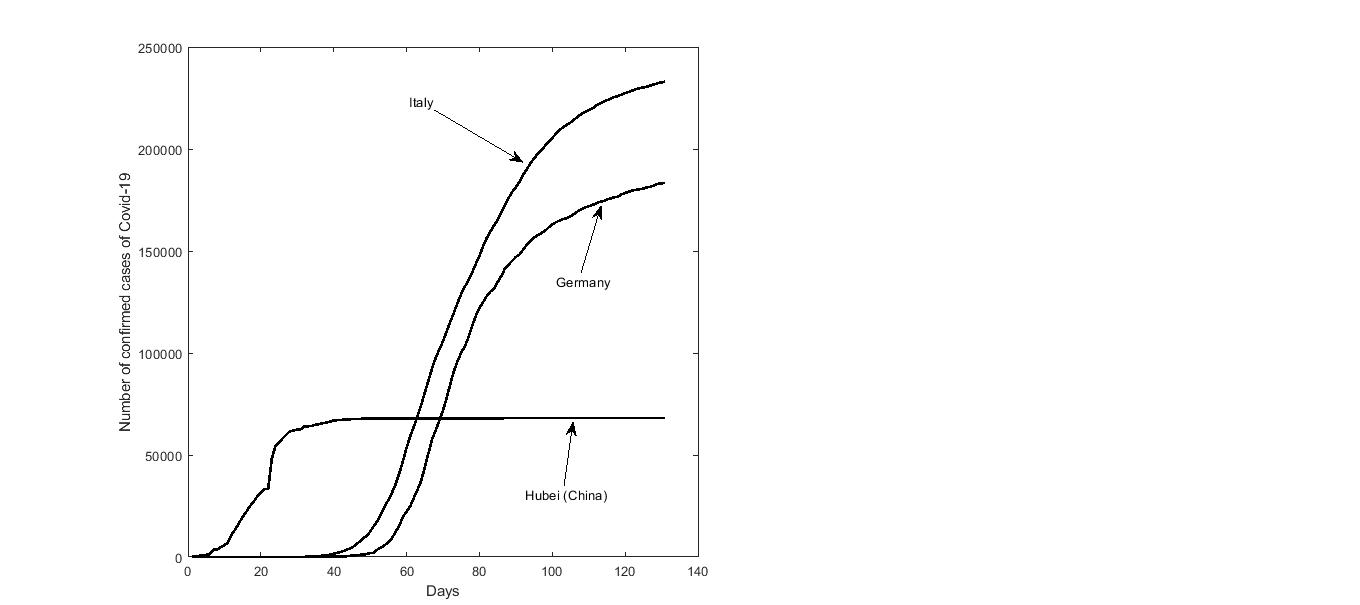}
	\caption{Confirmed cases of two countries (Italy and Germany) and Hubei (China) where the number of confirmed cases has almost reached to its peak.  From $22^{nd}$ January 2020 to $31^{st}$ May 2020 \cite{Johnhopdata}}
\end{figure}

Different approaches are used for modelling infectious diseases. Susceptible-Infectious-Recovered model and its extensions Susceptible-Infectious-Recovered-Deceased-Model, Susceptible-Exposed-Infectious-Recovered-model etc. have been applied to predict the spread of infectious diseases \cite{doi:10.1137/S0036144500371907}. The models are generally based on ordinary differential equations. These models can also be used to study the effect of public health interventions on the spread of infectious diseases. Recently, machine learning methods have  been applied to study modelling of infectious diseases \cite{CDCmachinelearning} .

In machine learning models are built using historical data and these models are used to predict the new outcome. Regression, classification, clustering, deep learning etc. \cite{Booksmachine,Goodfellow-et-al-2016} are some of the machine learning methods which have been successfully used in various domains such as image analysis, speech recognition, health informatics etc.   

Many applications of machine learning methods for Covid-19 have been proposed such as diagnosis and prognosis,  patient outcome prediction, tracking and predicting the outbreak, drug development, vaccine discovery, false news prediction, etc. \cite{Lancentml,review2,review5,bullock2020mapping,VAISHYA2020337}.
Many machine learning models have been used to model the number of confirmed cases of Covid-19 \cite{review2,review5,bullock2020mapping}. In this paper, we will review these machine learning models. There are other review papers on the related topics, we present the detailed analysis of these review papers and differentiate them from our paper. In this paper, we present a taxonomy to identify four broad
machine learning methodologies for predicting confirmed cases of Covid-19. Using this taxonomy, we present the comprehensive review of related published papers. We also present the challenges that have impacted this area. We further present suggestions to improve the performance of the machine learning models for the prediction of confirmed cases of Covid-19.

The paper is organized in following way. In the next section, we will discuss published reviews on the applications of machine learning methods for Covid-19. Section 3 will discuss various machine learning models to predict the number of confirmed cases of Covid-19. Challenges are presented in Section 4. The paper ends with suggestions to improve the accuracy of machine learning methods for the prediction of Covid-19 confirmed cases. 
    
\section{Related review papers}
A few review papers have been published that discuss applications of machine learning methods or artificial intelligence (AI) techniques for Covid-19. In this section, we will discuss these review papers. 
Naude \cite{review5} presents an early review of the applications of AI techniques for Covid-19. Prediction of the confirmed cases is one of the applications discussed in the paper. However, few related papers are discussed. Bullock et al. \cite{bullock2020mapping} discuss some of the studies that apply machine learning methods for handling Covid-19. They discuss some of the papers that apply machine learning methods to forecast the spread of Covid-19.
 Vaishya et al. \cite{VAISHYA2020337} discuss some of the applications of AI techniques for COVID-19 pandemic. They suggest that AI techniques can be used to predict the number of cases, however, no related papers are reviewed. Pham et al. \cite{review2}  review the applications of AI and big data for COVID-19 pandemic. In this paper, they review eight papers  related with prediction of spread of Covid-19.
 The study suggests that all the related review papers focus on many applications of machine learning methods for Covid-19. Prediction of the confirmed cases is not extensively covered in these papers. Our paper concentrates on only one application of machine learning methods for Covid-19. We try to include all the related papers majority of which are not covered in these review papers.
 
\section{Taxonomy for machine learning methodologies for the prediction of confirmed cases of Covid-19 }
The world is facing Covid-19 pandemic. The health policy makers need reliable estimates of the confirmed cases in future to make informed decisions about the required health facilities. They also want to estimate the effect of public health interventions to mitigate the epidemic. Machine learning methods have been applied to forecast the confirmed cases.
 
In this section, we will present a taxonomy to review the related published papers. The taxonomy identifies four research themes- \textit{traditional machine learning regression, deep learning regression, network analysis and social media and search queries data-based methods. }
There are some papers which may belong to more than one research themes, we take utmost care to place them in the most related research themes.

\subsection{Traditional machine learning regression} 
Regression analysis is a supervised machine learning approach which estimates the relationship between a dependent variable and independent variables. The independent variables are also called predictor variables \cite{Booksmachine}. These relationships are learned from the given data and are used to predict. Some of the regression analysis methods assume models and try to find the best parameters that fit the data to those models; for example multivariate linear regression (Eq. 1) it is assumed that the dependent variable $y$ is linearly dependent on $j$ predictor variables $x_i$ ($i$ is 1 to $j$). The task is to find out the values of $a_i$ so that the equation best fit the given data.

\begin{equation}
 y = \sum_{i=1}^{j}a_ix_i + a_0
\end{equation}

These relationships can be nonlinear. For example  polynomial regression of order more than one, S-shaped logistic curve (Eq. 2);

\begin{equation}
N_t = \frac{N_{max}}{1 + e^{-c(t-t_o)}}
\end{equation}

where $t$ is a predictor variable, $N_{max}$ is the maximum value of the curve, $t_{0}$ is the $t$ value of the sigmoid's midpoint, and $c$ is the logistic growth rate.
With the given data, the best values of  $N_{max}$, $t_{0}$ and $c$ are estimated. Then the equation is used to predict the number of curve, $N_t$, at given $t$. 

These methods work well if the assumed models are correct otherwise the performance may not be very accurate. Some of the machine learning methods do not assume any relationships and can learn complex relationships between predictor variables and dependent variable. Random forests \cite{breiman2001random} and neural networks \cite{Booksmachine} come under this category.

In the prediction of confirmed cases of Covid-19 using regression methods two approaches have been used.

\begin{enumerate}
	\item Time series analysis - In this approach the confirmed cases against the days curve is used for the prediction in the future. Two methodologies have been used for this purpose.
	\begin{itemize}
		\item $m$ previous days confirmed cases are used to predict the next day confirmed cases. A relationship is learned using multivariate regression analysis which is used to predict the confirmed cases in the future \cite{Gupta2020.04.01.20049825}.
		\item A relationship such as logistic curve is learned to predict the number of confirmed cases at a given day \cite{pavlyshenko2020regression,Maier742}.
	\end{itemize}

\item The relationships between confirmed cases of Covid-19 and the other factors such as temperature, humidity etc. are leaned from the data and the relationships are used to predict the number of confirmed cases with the new values of the factors  \cite{Pirouzbinaryclassifica}.
\end{enumerate}

We will present the detailed review of traditional machine learning regression methods for the confirmed cases of Covid-19. Gupta et al. \cite{Gupta2020.04.01.20049825} use polynomial regression to predict the number of confirmed cases in India. They use data from $30^{th}$ January till $25^{th}$ March 2020 as the training data and predict the number of COVID-19 patients in India for next two weeks.  Gu et al. \cite{GU2020} apply cubic regression equations which uses the number of days as the input variable to predict the conformed Covid 19 cases in China and world.

Pavlyshenko \cite{pavlyshenko2020regression} apply logistic curve to model COVID-19 spread. Model parameters are computed using Bayesian regression approach. The author argues that in the Bayesian inference, prior distributions can be set up by a Covid-19 expert which can be useful for small historical COVID19 datasets. Predictions are made for confirmed COVID19 cases in different countries. Batista \cite{batista2020.02.16.20023606} use logistic growth regression model to predict the the final size of the coronavirus epidemic. They predict the final value around 90000 which is way below the current value. Batista \cite{batista2020.03.11.20024901} does the similar calculation for final size of the second phase of the coronavirus using logistic model, the parameters of the logistic model are computed using regression analysis.  Ensembles are combinations of accurate and diverse models, they generally perform better than single model \cite{bookkuncheva}.  Buizza \cite{buizza2020weatherinspired} create an ensembles of  logistic curves to estimate the confirmed cases. These diverse logistic curves are generated by perturbing the training dataset. The model works well for China.

Petropoulos and Makridakis \cite{expo10.1371/journal.pone.0231236}  predict the global spread of Covid-19 using models from the exponential smoothing family. These have shown great prediction accuracy for short time series. Exponential smoothing models are useful to capture various types of trends. This approach is opposite to S-Curve (logistic curve) approach that assumes convergence. Stubinger and Schneider \cite{leadlag} argues that the Covid-19 spread in China first and then other countries got affected by it. Therefore, there are lead-lag effects between spreads of Covid-19 in different countries. The relationships can be exploited to predict the spread of Covid-19 in other countries using the  data of China. The relationships are computed using dynamic time warping. 

Tobias et al. \cite{TOBIAS2020138539} predict the number of the confirmed cases in the Italy and Spain under lockdown using quasi-Poisson regression model.  Interaction model is applied to compute the variations in confirmed cases trends.  Xu et al. \cite{Xu2020.04.08.20057943} apply a rolling growth curve approach (RGCA) to estimate the spread in USA. The model uses the number of daily hospitalized COVID-19 patients  as the independent variable. The authors argue that the number of confirmed deaths due to Covid-19 or the number of daily hospitalized COVID-19 patients is more reliable than the reported number of COVID-19 cases.  Li et al. \cite{Liregression2020.03.01.20029819} propose a regression equation to predict the early spread of Covid-19 in China. In this equation $log$ value of the summation of confirmed cases and 34 is linearly dependent on the days. An adaptive neuro-fuzzy inference system (ANFIS) is applied to forecast the number of conformed cases in China in  the time-series framework \cite{jcm9030674}. Enhanced flower pollination algorithm (FPA) and salp swarm algorithm (SSA) are used for selecting  parameters of ANFIS. The results suggest that the FPA and SSA perform better than the other parameter selection methods.

Maier and Brockmann \cite{Maier742} demonstrate that the scaling law, number of confirmed cases is proportional to  $(days)^x$, is universal for confirmed cases in affected provinces in Mainland China universal, with a range of exponents ($x$) = 2.1 $\pm$ 0.3. They suggests that effective containment strategies can be the reason for this behaviour. It is shown that all country-specific infection rates follow a power
law growth behaviour \cite{singer2020shortterm}. Different countries have different scaling exponents. The authors calculate scaling exponents for different countries. It is shown \cite{CASSARO2020138834} that cumulative distribution
function \cite{Zan2006} can be used to predict the spread of the Covid-19 using days as the predictor variable. 

Gupta and Pal \cite{Gupta2020.03.26.20044511} investigate the application of ARIMA (Auto-Regressive Integrated Moving Average) time series method for the prediction of confirmed covid 19 cases in India.  ARIMA model is used to estimate the spread in China, Italy, South Korea, Iran and Thailand \cite{Dehesh2020.03.13.20035345}. The model predicts stable trend in China and Thailand in future whereas the model predicts that Iran and Italy will have unstable trends in future. Using time-series framework, Chakraborty and Ghosh \cite{Chakraborty2020.04.09.20059311} combine ARIMA model and Wavelet-based forecasting model to  generate ten days ahead estimates of Covid-19 spread in various countries.

Perc et al. \cite{Perc10.3389/fphy.2020.00127} develop an iteration method to estimates the transmission of Covid-19 that requires the daily values of confirmed cases as input. It accounts for expected recoveries and deaths using a parameter. Covid-19 spread is predicted for various countries using the proposed model.

Lua et al. \cite{LAU2020} apply linear regression to suggest that countries with lower Healthcare Access and Quality (HAQ) Index may have larger number of  unreported cases of Covid-19.
Pirouz et al. \cite{Pirouzbinaryclassifica} uses a group method of data handling (GMDH) type of neural network \cite{Ivakhnenko1988SelfOrganizingMI} to investigate the relationship between environmental and urban factors  and the number of confirmed cases. It is also shown using 42 datasets in four countries, including China, Japan, South Korea, and Italy to show that there is a very low correlation between them, therefore different models should be created for these datasets.
Zhao et al. \cite{Zhao2020.02.07.20021196} propose that the daily traffic from Wuhan and the total traffic in this period can be used to estimate the spread in Chinese cities in January 2020. Multiple regression models are developed which use number of passengers and local population as predictor variables to explain the variance of the number of cases in the infected cities. Ensembles of ten different machine learning algorithms are used to estimate the spread of Covid-19 using climate variables (monthly mean temperature, interaction term
between monthly minimum temperature and maximum temperature, monthly precipitation sum, downward surface short-wave radiation, and actual
evapotranspiration) \cite{Araujo2020.03.12.20034728}.

Wang et al. \cite{wang2020high} apply linear regression framework to suggest that high temperature and high humidity significantly reduce the transmission of COVID-19. They use data from china in their experiments.  Oliveiros et al. \cite{Oliveiros2020.03.05.20031872}  apply a linear regression model to study the effect of temperature, humidity, precipitation, and wind speed on the doubling rate of COVID-19 spread in China. 

\subsection{Deep learning regression}
Deep learning is generally related with artificial neural networks which mimic human brain. Deep neural networks have large number of hidden layers \cite{Goodfellow-et-al-2016}. Deep neural networks have shown excellent performance in various domains image analysis, speech recognition, text analysis etc. Various types of deep learning neural networks have been applied to predict the spread of Covid-19. 

Long short-term memory (LSTM) is an artificial recurrent neural network architecture which is used to study a time series and predict the future events \cite{Goodfellow-et-al-2016}. Tomar and Gupta \cite{TOMAR2020138762} investigate the use of LSTM for the number of confirmed cases in India. The predicted  cases are very close to the official number of cases.
Hu et al. \cite{hu2020artificial} use  modified auto-encoders \cite{CHARTE201878} to model Covid-19 time series of confirmed cases in various Chinese cities. They use the trained model to predict six-step to ten-step forecasting. Yang et al. \cite{JTD36385} use LSTM to predict the spread in China in Feb. 2020. The LSTM is trained using the SERS 2003 data. COVID-19 epidemiological parameters such as transmission probability, incubation rate, etc. are incorporated in the model. The model predicts that the number of confirmed cases in China should peak by late February. Fong et al. \cite{smalldatasets} demonstrates that for small datasets  polynomial neural network with corrective feedback method outperform linear regression, support vector machines \cite{Booksmachine} and ARIMA methods. Deep learning-based Composite Monte-Carlo simulation \cite{FONG2020106282}  is used in conjunction of fuzzy rule induction techniques to predict the spread in China. Deep learning produces better fitted Monte Carlo outputs which lead to a better prediction. It is  demonstrated that the combination of LSTM and gated recurrent unit perform better than individual methods to predict the confirmed cases in the world \cite{Bandyopadhyay2020.03.25.20043505}. Nonlinear autoregressive artificial neural networks are used to predict the spread in many countries \cite{nonlinearauto}. Huang et al. \cite{Huang2020.03.23.20041608} demonstrate that convolutional neural networks \cite{Goodfellow-et-al-2016} outperform other deep learning models such as gated recurrent unit, LSTM and multilayer perceptron in the prediction of confirmed cases of Covid-19 in Chinese cities.

\subsection{Network analysis}
Networks or graphs consist of nodes and edges \cite{barabasi2016network}. Nodes represent the entities. Edges represent the connections between nodes. Analysis of 
graphs or networks is an important research area of machine learning. Web mining, social networks analysis, Community Detection etc. are some of the applications of the networks analysis \cite{barabasi2016network}.
 Humans are connected with other humans, therefore they can be represented by networks. These networks have been used to study the spread of the  infectious diseases \cite{barabasi2016network}. When one of nodes of a network gets infected, it can infect other nodes which are  connected to the infected node. This process continues and the disease spread to the other nodes of the network. In this section, we will discuss those papers that use network analysis to predict the number of confirmed cases of  Covid-19.

It has been observed that virological transmission usually satisfies the Gaussian distribution. Therefore, this  is used to predict the Covid-19 transmission \cite{LI2020282}, it is assumed that an infected person can infect 1 to $\infty$ persons. The model is used to estimate the transmission in China and the other countries. Zhuang et al. \cite{ZHUANG2020308} use a stochastic model to estimate the  confirmed cases in Republic of Korea and Italy. In this model, it is assumed that the number of secondary cases associated with a primary COVID-19 case follows a negative binomial distribution. The mean parameter represents the basic reproduction number of COVID-19 whereas the dispersion parameter represents the likelihood of occurrence of other factors that can effect spread like super-spreading events. 
The current growth closely follows power-law kinetics in China, indicative of an underlying fractal or small-world network (a small average shortest path length, and a large clustering coefficient \cite{barabasi2016network})of connections between susceptible and infected individuals \cite{Ziff2020.02.16.20023820}. Li et al \cite{li2020scaling} demonstrate that the
spread of Covid-19 closely follows a power-law kinetics in China during January 2020 to February 2020. It suggests that the underlying network has small-world property.

Herrmann and Schwartz \cite{Herrmann2020.04.02.20050468} propose that network of interactions can be used to estimate the spread of  Covid-19. They use scale-free networks (a scale-free network's degree distribution follow a power law) with Susceptible-Infected-Susceptible model \cite{PhysRevE.65.055103}, with the model parameters computed for COVID-19. Their results
show that directly targeting hubs in the network is far more effective than randomly decreasing the number of connections between individuals. 

It is shown that by transforming the time-series of COVID-19 infection curve to a visibility
graph one can study the time-series as a complex network \cite{Lacasa4972}. Complex-network-based splines regression method is proposed for the prediction of confirmed cases in Greece \cite{demertzis2020modeling}. The proposed method outperforms both the cubic regression model and the randomly-calibrated splines
regression model.

Pujari and Shekatkar \cite{Pujari2020.03.13.20035386} propose a hybrid model to predict the spread of Covid-19 in Indian cities.
Susceptible-Infectious-Recovered models are created for individual cities. The migration among cities are modelled using transportation networks. Indian aviation and railway networks are used as transportation networks. Biswas and Sen \cite{biswas2020spacetime} use Susceptible-Infected-Removed model of epidemic spreading on Euclidean
networks to study  space-time dependence of Covid-19 spread in many countries. It is assumed that the disease can be transmitted to a nearest neighbour and to some random other agent who is connected with a probability decaying algebraically with the Euclidean distance separating them.

Gross et al. \cite{Gross2020} study the spatial dynamics of the COVID-19 in Hubei and other provinces of China. It is demonstrated that power laws hold for the
number of confirmed cases in each province as a function
of the province population and the distance from
Hubei.

Hybrid non-linear cellular automata (HNLCA) classifier is trained on different parameters such as movement of the people, various transmission rates, vulnerable people in the region, etc. to estimate the spread in India \cite{PokkuluriDeviNedunuri2020}. 
Regression analysis is carried out to evaluate the relationship between migration and the spread of Covid-19 in China \cite{Chinamigrant}. Bivariate correlation analysis is used to extract the strength of these relationships in various cities of China. 
\subsection{Social media and search queries data-based methods} 
Internet search queries and social media have emerged as rich sources of data. This data is used to predict and monitor infectious diseases \cite{naturecommunication}.  It is easy to collect this data, however there is a problem of noisy data. Internet search queries and social media have been used to predict the number of confirmed cases of Covid-19, we will review these research works in this section.

Qin et al. \cite{Quinsocialijerph17072365} predict the number of Covid-19 cases by using  social media
search indexes (SMSI) of Covid-19 symptoms (dry cough, fever, chest distress, coronavirus, and pneumonia) collected from Baidu search engine data as the independent variables. Five regression methods, subset selection, forward selection, lasso regression, ridge regression, and elastic net, are used to determine the relationships between the number of Covid-19 cases and independent variables. The subset selection method produces the best result. Jahanbin and  Rahmanian \cite{twitterIranauthor}  conclude that tweets extracted from twitter can be used to estimate the spread of Covid-19. Fuzzy rule-based evolutionary algorithm called Eclass1-MIMO is used to model the spread. The results suggest that geographical origins of tweets posted about COVID-19 are consistent with the number of confirmed cases of Covid-19.

ARGOnet \cite{naturecommunication} combines  AutoRegression with General Online information (ARGO) with  spatio-temporal information about influenza activity to predict influenza transmission. Liu et al. \cite{liu2020machine} apply AGROnet to predict Covid-19 cases, the  model uses the data from  official health reports from \textit{Chinese Center Disease for Control and Prevention}, Covid-19 related internet search activity from Baidu, media activity reported by Media Cloud, and daily forecasts of COVID-19 activity by an agent-based mechanistic model. They apply clustering to get spatio-temporal Covid-19 information across Chinese provinces required for AGROnet. ARGOnet outperforms an autoregressive model trained only on historical confirmed cases.

Ayyoubzadeh et al. \cite{info:doi/10.2196/18828} use search statistics collected from  Google Trends for search queries related to COVID-19 such as Corona, COVID-19, Coronavirus, hand washing, Antiseptic, etc. to predict the  number of confirmed cases in the next days in Iran. Linear regression and LSTM models are used for the prediction. Linear regression model performs better.

Lampos et al. \cite{lampos2020tracking} use the data pertaining to google search queries  and basic news media coverage metric associated with Covid-19. Elastic net models are trained using the data. The authors also study the transferability of their models between
countries. It is proposed that Twitter sentiment analysis can be used to predict the spread of COVID19 Outbreak \cite{twitterdubey}. 

\section{Challenges}
Many machine learning methods have been used to predict the number of confirmed cases of Covid-19. However, there are many challenges for the accurate prediction by machine learning methods. In this section, we will discuss these challanges.

Time series framework is very popular in prediction of the confirmed cases. In many countries the first case of Covid-19 came in January 2020 or after.  Therefore till 30th May 2020, the number of  data points is less 150. It is difficult to train accurate machine learning models with such small datasets. Deep learning methods are successful because of large training data which is not available for Covid-19 confirmed cases prediction task. It is difficult to select proper architectures and parameters for deep learning neural networks with small datasets. 

The lack of historical data is a major problem. Pandemics are rare and the characteristics of Covid-19 are different than other coronaviruses such as SARS and MERS\cite{PETROSILLO2020729}. Therefore their data cannot be used for Covid-19. 

It is argued that many countries are not doing enough testing \cite{covidtesting}. Therefore, it is impossible to have correct number of confirmed cases in these countries. Training machine learning algorithms with datasets of poor quality will lead to misleading conclusions.

Governments take different preventive steps such as lockdown, social distancing etc. to slow the spread of Covid-19. The effectiveness of these measures are different in countries. Inclusion of the effects of these measures in machine learning methods is a challenging task.

It has been estimated that about 80$\%$ of people with COVID-19 are
mild or asymptomatic cases \cite{mild}. Many of these people do not get tested, therefore the numbers of confirmed cases in countries are not accurate. However, these people contribute to the new cases. It makes creating accurate machine learning models for Covid-19 confirmed cases a difficult task.

Social stigma attached to Covid-19 \cite{Scialstigma} in many countries force suspected Covid-19 patients to stay away from medical facilities. Therefore, many confirmed cases are not recorded. However, they contribute to new confirmed cases.

Logistic curve regression have been applied to predict the number of the confirmed caseas. It is shown that by using logistic curve regression, a real-world epidemic outbreak can be predicted reliably only in the short term \cite{Logisticno}. Therefore, logistic curve regression may not produce accurate prediction for long term.

Social media data has been applied for the prediction of Covid-19 cases. This data is huge, however, the data is noisy.  Search queries have been used to predict the number of confirmed cases. However, many normal people also try to find the information about Covid-19. Therefore, it is difficult to predict the confirmed cases only on the basis of search queries. Furthermore, there is a vast difference in Internet penetration is various countries. The number of Internet users are vastly different in countries \cite{internetusers}.
The amount of social media data and search queries data is quite different in countries.Therefore, it is difficult to find prediction models which may work for many countries.

Predicting the number of confirmed cases of Covid-19 using network analysis requires accurate estimation of the transmission of Covid-19 from one person to another person. As Covid-19 is a new pandemic, all the characteristics of this pandemic are not known. Hence, the predicted confirmed cases may not be accurate.

Structures of networks also play an important roles in accurate prediction. In a country the connections between nodes are dependent on the culture, environment, population density etc. Some of these factors such as culture are difficult to quantify. Therefore, it is difficult to estimate the connections between nodes accurately. Furthermore, models in one country cannot be applied to other countries easily as they have different factors.   

\section{Suggestions}
We present some suggestions that can be used to address the challenges for machine learning methods for the prediction of confirmed cases of Covid-19. 

Machine learning methods have not been very successful because of the challenges discussed in the last section. Established epidemiological models, such as Susceptible-Infectious-Recovered models, have been successfully used for modelling infectious diseases. The hybrid models of  machine learning algorithms and epidemiological models will be a promising research area.
	
	Machine learning experts should work with epidemiologists to understand the data well and to select the parameters of  machine learning methods for the prediction of confirmed cases of Covid-19. 
	
	As it is difficult to predict the number of confirmed cases of Covid-19 accurately from one type of data. Models trained on different types of data may be combined to predict the number of confirmed cases. 
	
Countries are at different stages of the pandemic. Therefore, data from one country that are ahead in the epidemic curve  may be used  for other countries in earlier stages of the epidemic curve. There have been some attempts in this research direction \cite{leadlag}. 
	Transfer learning deals with training a model for  a given problem and using it to related but different problem \cite{DBLP:conf/www/ZouLC19}. Transfer learning has been applied for the prediction of confirmed cases by using the model trained on data from one country in which is ahead in the epidemic curve to other countries still in earlier stages of the epidemic curve \cite{lampos2020tracking}. More research is required in this direction.
	  
Some countries have better facilities of collecting data. This data can be used for countries with less facilities of collecting data. The data should be modified for the local context for better representation of a country.
 
 In countries where social stigma attached with Covid-19 scare people from taking medical help,  new sources for data such as medicine buying pattern, online-purchasing pattern, shopping pattern etc. should be investigated  for the prediction of covid-19 cases.
 
 In this paper, we identified four major research themes to predict the number of confirmed cases of Covid-19. We presented a comprehensive state-of-the-art review of the related research papers within them. We discussed the challenges in this research area and presented suggestions to address them. We believe that the review paper will be helpful to the researchers to develop an in-depth understanding of the research area. This paper may lead to generate novel ideas of using machine learning methods for accurately predicting the number of confirmed cases of Covid-19. This will be very beneficial to mankind.

\bibliography{mybibfile}

\begin{thebibliography}{10}
\expandafter\ifx\csname url\endcsname\relax
  \def\url#1{\texttt{#1}}\fi
\expandafter\ifx\csname urlprefix\endcsname\relax\def\urlprefix{URL }\fi
\expandafter\ifx\csname href\endcsname\relax
  \def\href#1#2{#2} \def\path#1{#1}\fi

\bibitem{Whocovid}
Naming the coronavirus disease (covid-19) and the virus that causes it. {W}orld
  {H}ealth {O}rganization.,
  \url{https://www.who.int/emergencies/diseases/novel-coronavirus-2019/technical-guidance/naming-the-coronavirus-disease-(covid-2019)-and-the-virus-that-causes-it},
  accessed $26^{th}$ May 2020.

\bibitem{Whochina}
Novel coronavirus in {C}hina. {W}orld {H}ealth {O}rganization.,
  \url{https://www.who.int/csr/don/12-january-2020-novel-coronavirus-china/en/},
  accessed $26^{th}$ May 2020.

\bibitem{Whopandemic}
{WHO} {Director-General's} opening remarks at the media briefing on {C}ovid-19.
  {W}orld {H}ealth {O}rganization (press release). 11 march 2020,
  \url{https://www.who.int/dg/speeches/detail/who-director-general-s-opening-remarks-at-the-media-briefing-on-covid-19---11-march-2020},
  accessed $26^{th}$ May 2020.

\bibitem{Johnhop}
Covid-19 dashboard by the center for systems science and engineering (csse) at
  {J}ohns {H}opkins {U}niversity, {USA},
  \url{https://gisanddata.maps.arcgis.com/apps/opsdashboard/index.html#/bda7594740fd40299423467b48e9ecf6},
  accessed 26th May 2020.

\bibitem{Johnhopdata}
Novel coronavirus (covid-19) cases, provided by {J}ohns {H}opkins {U}niversity,
  {USA}, \url{https://github.com/CSSEGISandData/COVID-19}, accessed $31^{th}$
  May 2020.

\bibitem{doi:10.1137/S0036144500371907}
H.~W. Hethcote, The mathematics of infectious diseases, SIAM Review 42~(4)
  (2000) 599--653.
\newblock \href {http://dx.doi.org/10.1137/S0036144500371907}
  {\path{doi:10.1137/S0036144500371907}}.

\bibitem{CDCmachinelearning}
K.~Hao, This is how the cdc is trying to forecast coronavirus spread,
  \url{https://www.technologyreview.com/2020/03/13/905313/cdc-cmu-forecasts-coronavirus-spread/},
  accessed $25^{th}$ May 2020.

\bibitem{Booksmachine}
C.~M. Bishop, Pattern {R}ecognition and {M}achine {L}earning, Springer-Verlag
  New York Inc, 2008.

\bibitem{Goodfellow-et-al-2016}
I.~Goodfellow, Y.~Bengio, A.~Courville, Deep Learning, MIT Press, 2016,
  \url{http://www.deeplearningbook.org}.

\bibitem{Lancentml}
B.~McCall, Covid-19 and artificial intelligence: protecting health-care workers
  and curbing the spread, The Lancet Digital Health 2~(4) (2020) e166 -- e167.

\bibitem{review2}
Q.~Pham, D.~C. Nguyen, T.~Huynh-The, W.~Hwang, P.~Pathirana, Artificial
  intelligence (ai) and big data for coronavirus (covid-19) pandemic: A survey
  on the state-of-the-arts,
  \url{https://www.preprints.org/manuscript/202004.0383/v1}.
\newblock \href {http://arxiv.org/abs/Preprints 2020, 2020040383}
  {\path{arXiv:Preprints 2020, 2020040383}}, \href {http://dx.doi.org/doi:
  10.20944/preprints202004.0383.v1} {\path{doi:doi:
  10.20944/preprints202004.0383.v1}}.

\bibitem{review5}
Artificial intelligence against covid-19: An early review,
  \url{https://www.iza.org/publications/dp/13110/artificial-intelligence-against-covid-19-an-early-reviewBatis}.
\newblock \href {http://arxiv.org/abs/IZA Discussion Paper No. 13110}
  {\path{arXiv:IZA Discussion Paper No. 13110}}.

\bibitem{bullock2020mapping}
J.~Bullock, A.~Luccioni, K.~H. Pham, C.~S.~N. Lam, M.~Luengo-Oroz, Mapping the
  landscape of artificial intelligence applications against covid-19 (2020).
\newblock \href {http://arxiv.org/abs/2003.11336} {\path{arXiv:2003.11336}}.

\bibitem{VAISHYA2020337}
R.~Vaishya, M.~Javaid, I.~H. Khan, A.~Haleem,
  \href{http://www.sciencedirect.com/science/article/pii/S1871402120300771}{Artificial
  intelligence (ai) applications for covid-19 pandemic}, Diabetes and Metabolic
  Syndrome Clinical Research and Reviews 14~(4) (2020) 337 -- 339.
\newblock \href {http://dx.doi.org/https://doi.org/10.1016/j.dsx.2020.04.012}
  {\path{doi:https://doi.org/10.1016/j.dsx.2020.04.012}}.
\newline\urlprefix\url{http://www.sciencedirect.com/science/article/pii/S1871402120300771}

\bibitem{breiman2001random}
L.~Breiman, Random forests, Machine Learning 45~(1) (2001) 5--32.
\newblock \href {http://dx.doi.org/10.1023/A:1010933404324}
  {\path{doi:10.1023/A:1010933404324}}.

\bibitem{Gupta2020.04.01.20049825}
R.~Gupta, G.~Pandey, P.~Chaudhary, S.~K. Pal,
  \href{https://www.medrxiv.org/content/early/2020/04/03/2020.04.01.20049825}{Seir
  and regression model based covid-19 outbreak predictions in india},
  medRxiv\href
  {http://arxiv.org/abs/https://www.medrxiv.org/content/early/2020/04/03/2020.04.01.20049825.full.pdf}
  {\path{arXiv:https://www.medrxiv.org/content/early/2020/04/03/2020.04.01.20049825.full.pdf}},
  \href {http://dx.doi.org/10.1101/2020.04.01.20049825}
  {\path{doi:10.1101/2020.04.01.20049825}}.
\newline\urlprefix\url{https://www.medrxiv.org/content/early/2020/04/03/2020.04.01.20049825}

\bibitem{pavlyshenko2020regression}
B.~M. Pavlyshenko, Regression approach for modeling covid-19 spread and its
  impact on stock market (2020).
\newblock \href {http://arxiv.org/abs/2004.01489} {\path{arXiv:2004.01489}}.

\bibitem{Maier742}
B.~F. Maier, D.~Brockmann,
  \href{https://science.sciencemag.org/content/368/6492/742}{Effective
  containment explains subexponential growth in recent confirmed covid-19 cases
  in china}, Science 368~(6492) (2020) 742--746.
\newblock \href
  {http://arxiv.org/abs/https://science.sciencemag.org/content/368/6492/742.full.pdf}
  {\path{arXiv:https://science.sciencemag.org/content/368/6492/742.full.pdf}},
  \href {http://dx.doi.org/10.1126/science.abb4557}
  {\path{doi:10.1126/science.abb4557}}.
\newline\urlprefix\url{https://science.sciencemag.org/content/368/6492/742}

\bibitem{Pirouzbinaryclassifica}
B.~Pirouz, S.~S. Haghshenas, S.~S. Haghshenas, P.~Piro, Investigating a serious
  challenge in the sustainable development process: Analysis of confirmed cases
  of covid-19 (new type of coronavirus) through a binary classification using
  artificial intelligence and regression analysis, Sustainability 12~(6) (2020)
  2427.

\bibitem{GU2020}
C.~Gu, J.~Zhu, Y.~Sun, K.~Zhou, J.~Gu,
  \href{http://www.sciencedirect.com/science/article/pii/S2095927320301122}{The
  inflection point about covid-19 may have passed}, Science Bulletin\href
  {http://dx.doi.org/https://doi.org/10.1016/j.scib.2020.02.025}
  {\path{doi:https://doi.org/10.1016/j.scib.2020.02.025}}.
\newline\urlprefix\url{http://www.sciencedirect.com/science/article/pii/S2095927320301122}

\bibitem{batista2020.02.16.20023606}
M.~Batista, Estimation of the final size of the covid-19 epidemic, medRxiv\href
  {http://dx.doi.org/10.1101/2020.02.16.20023606}
  {\path{doi:10.1101/2020.02.16.20023606}}.

\bibitem{batista2020.03.11.20024901}
M.~Batista,
  \href{https://www.medrxiv.org/content/early/2020/03/17/2020.03.11.20024901}{Estimation
  of the final size of the second phase of the coronavirus epidemic by the
  logistic model}, medRxiv\href
  {http://arxiv.org/abs/https://www.medrxiv.org/content/early/2020/03/17/2020.03.11.20024901.full.pdf}
  {\path{arXiv:https://www.medrxiv.org/content/early/2020/03/17/2020.03.11.20024901.full.pdf}},
  \href {http://dx.doi.org/10.1101/2020.03.11.20024901}
  {\path{doi:10.1101/2020.03.11.20024901}}.
\newline\urlprefix\url{https://www.medrxiv.org/content/early/2020/03/17/2020.03.11.20024901}

\bibitem{bookkuncheva}
L.~I. Kuncheva, Combining Pattern Classifiers: Methods and Algorithms,
  Wiley-Interscience, USA, 2004.

\bibitem{buizza2020weatherinspired}
R.~Buizza, Weather-inspired ensemble-based probabilistic prediction of covid-19
  (2020).
\newblock \href {http://arxiv.org/abs/2003.06418} {\path{arXiv:2003.06418}}.

\bibitem{expo10.1371/journal.pone.0231236}
F.~Petropoulos, S.~Makridakis,
  \href{https://doi.org/10.1371/journal.pone.0231236}{Forecasting the novel
  coronavirus covid-19}, PLOS ONE 15~(3) (2020) 1--8.
\newblock \href {http://dx.doi.org/10.1371/journal.pone.0231236}
  {\path{doi:10.1371/journal.pone.0231236}}.
\newline\urlprefix\url{https://doi.org/10.1371/journal.pone.0231236}

\bibitem{leadlag}
J.~Stubinger, L.~Schneider, Epidemiology of coronavirus covid-19: Forecasting
  the future incidence in different countries, Healthcare 99~(8(2)).

\bibitem{TOBIAS2020138539}
A.~Tobias, Evaluation of the lockdowns for the sars cov 2 epidemic in italy and
  spain after one month follow up, Science of The Total Environment 725 (2020)
  138539.

\bibitem{Xu2020.04.08.20057943}
S.~Xu, C.~Clarke, S.~Shetterly, K.~Narwaney,
  \href{https://www.medrxiv.org/content/early/2020/04/20/2020.04.08.20057943}{Estimating
  the growth rate and doubling time for short-term prediction and monitoring
  trend during the covid-19 pandemic with a sas macro}, medRxiv\href
  {http://arxiv.org/abs/https://www.medrxiv.org/content/early/2020/04/20/2020.04.08.20057943.full.pdf}
  {\path{arXiv:https://www.medrxiv.org/content/early/2020/04/20/2020.04.08.20057943.full.pdf}},
  \href {http://dx.doi.org/10.1101/2020.04.08.20057943}
  {\path{doi:10.1101/2020.04.08.20057943}}.
\newline\urlprefix\url{https://www.medrxiv.org/content/early/2020/04/20/2020.04.08.20057943}

\bibitem{Liregression2020.03.01.20029819}
Y.~Li, M.~Liang, X.~Yin, X.~Liu, M.~Hao, Z.~Hu, Y.~Wang, L.~Jin,
  \href{https://www.medrxiv.org/content/early/2020/03/05/2020.03.01.20029819}{Covid-19
  epidemic outside china: 34 founders and exponential growth}, medRxiv\href
  {http://arxiv.org/abs/https://www.medrxiv.org/content/early/2020/03/05/2020.03.01.20029819.full.pdf}
  {\path{arXiv:https://www.medrxiv.org/content/early/2020/03/05/2020.03.01.20029819.full.pdf}},
  \href {http://dx.doi.org/10.1101/2020.03.01.20029819}
  {\path{doi:10.1101/2020.03.01.20029819}}.
\newline\urlprefix\url{https://www.medrxiv.org/content/early/2020/03/05/2020.03.01.20029819}

\bibitem{jcm9030674}
M.~A. A.~A. qaness, A.~A. Ewees, H.~Fan, M.~A.~E. Aziz, Optimization method for
  forecasting confirmed cases of covid-19 in china, Journal of Clinical
  Medicine 9~(3).

\bibitem{singer2020shortterm}
H.~M. Singer, Short-term predictions of country-specific covid-19 infection
  rates based on power law scaling exponents (2020).
\newblock \href {http://arxiv.org/abs/2003.11997} {\path{arXiv:2003.11997}}.

\bibitem{CASSARO2020138834}
F.~A. Cassaro, L.~F. Pires,
  \href{http://www.sciencedirect.com/science/article/pii/S0048969720323512}{Can
  we predict the occurrence of covid-19 cases? considerations using a simple
  model of growth}, Science of The Total Environment 728 (2020) 138834.
\newblock \href
  {http://dx.doi.org/https://doi.org/10.1016/j.scitotenv.2020.138834}
  {\path{doi:https://doi.org/10.1016/j.scitotenv.2020.138834}}.
\newline\urlprefix\url{http://www.sciencedirect.com/science/article/pii/S0048969720323512}

\bibitem{Zan2006}
P.~Zandbergen, J.~Chakraborty, Improving environmental exposure analysis using
  cumulative distribution functions and individual geocoding, International
  Journal of Health Geographics 5~(23).

\bibitem{Gupta2020.03.26.20044511}
R.~Gupta, S.~K. Pal,
  \href{https://www.medrxiv.org/content/early/2020/03/30/2020.03.26.20044511}{Trend
  analysis and forecasting of covid-19 outbreak in india}, medRxiv\href
  {http://arxiv.org/abs/https://www.medrxiv.org/content/early/2020/03/30/2020.03.26.20044511.full.pdf}
  {\path{arXiv:https://www.medrxiv.org/content/early/2020/03/30/2020.03.26.20044511.full.pdf}},
  \href {http://dx.doi.org/10.1101/2020.03.26.20044511}
  {\path{doi:10.1101/2020.03.26.20044511}}.
\newline\urlprefix\url{https://www.medrxiv.org/content/early/2020/03/30/2020.03.26.20044511}

\bibitem{Dehesh2020.03.13.20035345}
T.~Dehesh, H.~Fard, P.~Dehesh,
  \href{https://www.medrxiv.org/content/early/2020/03/18/2020.03.13.20035345}{Forecasting
  of covid-19 confirmed cases in different countries with arima models},
  medRxiv\href
  {http://arxiv.org/abs/https://www.medrxiv.org/content/early/2020/03/18/2020.03.13.20035345.full.pdf}
  {\path{arXiv:https://www.medrxiv.org/content/early/2020/03/18/2020.03.13.20035345.full.pdf}},
  \href {http://dx.doi.org/10.1101/2020.03.13.20035345}
  {\path{doi:10.1101/2020.03.13.20035345}}.
\newline\urlprefix\url{https://www.medrxiv.org/content/early/2020/03/18/2020.03.13.20035345}

\bibitem{Chakraborty2020.04.09.20059311}
T.~Chakraborty, I.~Ghosh,
  \href{https://www.medrxiv.org/content/early/2020/04/14/2020.04.09.20059311}{Real-time
  forecasts and risk assessment of novel coronavirus (covid-19) cases: A
  data-driven analysis}, medRxiv\href
  {http://arxiv.org/abs/https://www.medrxiv.org/content/early/2020/04/14/2020.04.09.20059311.full.pdf}
  {\path{arXiv:https://www.medrxiv.org/content/early/2020/04/14/2020.04.09.20059311.full.pdf}},
  \href {http://dx.doi.org/10.1101/2020.04.09.20059311}
  {\path{doi:10.1101/2020.04.09.20059311}}.
\newline\urlprefix\url{https://www.medrxiv.org/content/early/2020/04/14/2020.04.09.20059311}

\bibitem{Perc10.3389/fphy.2020.00127}
M.~Perc, N.~G. Miksic, M.~Slavinec, A.~Stozer,
  \href{https://www.frontiersin.org/article/10.3389/fphy.2020.00127}{Forecasting
  covid-19}, Frontiers in Physics 8 (2020) 127.
\newblock \href {http://dx.doi.org/10.3389/fphy.2020.00127}
  {\path{doi:10.3389/fphy.2020.00127}}.
\newline\urlprefix\url{https://www.frontiersin.org/article/10.3389/fphy.2020.00127}

\bibitem{LAU2020}
H.~Lau, V.~Khosrawipour, P.~Kocbach, A.~Mikolajczyk, H.~Ichii, J.~Schubert,
  J.~Bania, T.~Khosrawipour,
  \href{http://www.sciencedirect.com/science/article/pii/S1684118220300736}{Internationally
  lost covid-19 cases}, Journal of Microbiology, Immunology and Infection\href
  {http://dx.doi.org/https://doi.org/10.1016/j.jmii.2020.03.013}
  {\path{doi:https://doi.org/10.1016/j.jmii.2020.03.013}}.
\newline\urlprefix\url{http://www.sciencedirect.com/science/article/pii/S1684118220300736}

\bibitem{Ivakhnenko1988SelfOrganizingMI}
A.~G. Ivakhnenko, Self-organizing methods in modelling and clustering: Gmdh
  type algorithms, in: Systems Analysis and Simulation I, 1988, pp. 86--88.

\bibitem{Zhao2020.02.07.20021196}
X.~Zhao, X.~Liu, X.~Li,
  \href{https://www.medrxiv.org/content/early/2020/02/11/2020.02.07.20021196}{Tracking
  the spread of novel coronavirus (2019-ncov) based on big data}, medRxiv\href
  {http://arxiv.org/abs/https://www.medrxiv.org/content/early/2020/02/11/2020.02.07.20021196.full.pdf}
  {\path{arXiv:https://www.medrxiv.org/content/early/2020/02/11/2020.02.07.20021196.full.pdf}},
  \href {http://dx.doi.org/10.1101/2020.02.07.20021196}
  {\path{doi:10.1101/2020.02.07.20021196}}.
\newline\urlprefix\url{https://www.medrxiv.org/content/early/2020/02/11/2020.02.07.20021196}

\bibitem{Araujo2020.03.12.20034728}
M.~B. Araujo, B.~Naimi,
  \href{https://www.medrxiv.org/content/early/2020/04/07/2020.03.12.20034728}{Spread
  of sars-cov-2 coronavirus likely to be constrained by climate}, medRxiv\href
  {http://arxiv.org/abs/https://www.medrxiv.org/content/early/2020/04/07/2020.03.12.20034728.full.pdf}
  {\path{arXiv:https://www.medrxiv.org/content/early/2020/04/07/2020.03.12.20034728.full.pdf}},
  \href {http://dx.doi.org/10.1101/2020.03.12.20034728}
  {\path{doi:10.1101/2020.03.12.20034728}}.
\newline\urlprefix\url{https://www.medrxiv.org/content/early/2020/04/07/2020.03.12.20034728}

\bibitem{wang2020high}
J.~Wang, K.~Tang, K.~Feng, W.~Lv, High temperature and high humidity reduce the
  transmission of covid-19 (2020).
\newblock \href {http://arxiv.org/abs/2003.05003} {\path{arXiv:2003.05003}}.

\bibitem{Oliveiros2020.03.05.20031872}
B.~Oliveiros, L.~Caramelo, N.~C. Ferreira, F.~Caramelo,
  \href{https://www.medrxiv.org/content/early/2020/03/08/2020.03.05.20031872}{Role
  of temperature and humidity in the modulation of the doubling time of
  covid-19 cases}, medRxiv\href
  {http://arxiv.org/abs/https://www.medrxiv.org/content/early/2020/03/08/2020.03.05.20031872.full.pdf}
  {\path{arXiv:https://www.medrxiv.org/content/early/2020/03/08/2020.03.05.20031872.full.pdf}},
  \href {http://dx.doi.org/10.1101/2020.03.05.20031872}
  {\path{doi:10.1101/2020.03.05.20031872}}.
\newline\urlprefix\url{https://www.medrxiv.org/content/early/2020/03/08/2020.03.05.20031872}

\bibitem{TOMAR2020138762}
A.~Tomar, N.~Gupta,
  \href{http://www.sciencedirect.com/science/article/pii/S0048969720322798}{Prediction
  for the spread of covid-19 in india and effectiveness of preventive
  measures}, Science of The Total Environment 728 (2020) 138762.
\newblock \href
  {http://dx.doi.org/https://doi.org/10.1016/j.scitotenv.2020.138762}
  {\path{doi:https://doi.org/10.1016/j.scitotenv.2020.138762}}.
\newline\urlprefix\url{http://www.sciencedirect.com/science/article/pii/S0048969720322798}

\bibitem{hu2020artificial}
Z.~Hu, Q.~Ge, S.~Li, L.~Jin, M.~Xiong, Artificial intelligence forecasting of
  covid-19 in china (2020).
\newblock \href {http://arxiv.org/abs/2002.07112} {\path{arXiv:2002.07112}}.

\bibitem{CHARTE201878}
D.~Charte, F.~Charte, S.~Garcia, M.~J.~D. Jesus, F.~Herrera,
  \href{http://www.sciencedirect.com/science/article/pii/S1566253517307844}{A
  practical tutorial on autoencoders for nonlinear feature fusion: Taxonomy,
  models, software and guidelines}, Information Fusion 44 (2018) 78 -- 96.
\newblock \href
  {http://dx.doi.org/https://doi.org/10.1016/j.inffus.2017.12.007}
  {\path{doi:https://doi.org/10.1016/j.inffus.2017.12.007}}.
\newline\urlprefix\url{http://www.sciencedirect.com/science/article/pii/S1566253517307844}

\bibitem{JTD36385}
Z.~Yang, Z.~Zeng, K.~Wang, S.-S. Wong, W.~Liang, M.~Zanin, P.~Liu, X.~Cao,
  Z.~Gao, Z.~Mai, J.~Liang, X.~Liu, S.~Li, Y.~Li, F.~Ye, W.~Guan, Y.~Yang,
  F.~Li, S.~Luo, Y.~Xie, B.~Liu, Z.~Wang, S.~Zhang, Y.~Wang, N.~Zhong, J.~He,
  Modified seir and ai prediction of the epidemics trend of covid-19 in china
  under public health interventions, Journal of Thoracic Disease 12~(3).

\bibitem{smalldatasets}
S.~J. Fong, N.~D. G.~Li, R.~G. Crespo, E.~Herrera-Viedma, Finding an accurate
  early forecasting model from small dataset: A case of 2019-ncov novel
  coronavirus outbreak, International Journal of Interactive Multimedia and
  Artificial Intelligence 6 (2020) 132--139.

\bibitem{FONG2020106282}
S.~J. Fong, G.~Li, N.~Dey, R.~G. Crespo, E.~Herrera-Viedma,
  \href{http://www.sciencedirect.com/science/article/pii/S1568494620302222}{Composite
  monte carlo decision making under high uncertainty of novel coronavirus
  epidemic using hybridized deep learning and fuzzy rule induction}, Applied
  Soft Computing 93 (2020) 106282.
\newblock \href {http://dx.doi.org/https://doi.org/10.1016/j.asoc.2020.106282}
  {\path{doi:https://doi.org/10.1016/j.asoc.2020.106282}}.
\newline\urlprefix\url{http://www.sciencedirect.com/science/article/pii/S1568494620302222}

\bibitem{Bandyopadhyay2020.03.25.20043505}
S.~K. Bandyopadhyay, S.~Dutta,
  \href{https://www.medrxiv.org/content/early/2020/03/30/2020.03.25.20043505}{Machine
  learning approach for confirmation of covid-19 cases: Positive, negative,
  death and release}, medRxiv\href
  {http://arxiv.org/abs/https://www.medrxiv.org/content/early/2020/03/30/2020.03.25.20043505.full.pdf}
  {\path{arXiv:https://www.medrxiv.org/content/early/2020/03/30/2020.03.25.20043505.full.pdf}},
  \href {http://dx.doi.org/10.1101/2020.03.25.20043505}
  {\path{doi:10.1101/2020.03.25.20043505}}.
\newline\urlprefix\url{https://www.medrxiv.org/content/early/2020/03/30/2020.03.25.20043505}

\bibitem{nonlinearauto}
N.~M. Ghazaly, M.~A. Abdel-Fattah, A.~A.~A. El-Aziz, Novel coronavirus
  forecasting model using nonlinear autoregressive artificial neural network,
  International Journal of Advanced Science and Technology 29~(5s).

\bibitem{Huang2020.03.23.20041608}
C.-J. Huang, Y.-H. Chen, Y.~Ma, P.-H. Kuo,
  \href{https://www.medrxiv.org/content/early/2020/03/27/2020.03.23.20041608}{Multiple-input
  deep convolutional neural network model for covid-19 forecasting in china},
  medRxiv\href
  {http://arxiv.org/abs/https://www.medrxiv.org/content/early/2020/03/27/2020.03.23.20041608.full.pdf}
  {\path{arXiv:https://www.medrxiv.org/content/early/2020/03/27/2020.03.23.20041608.full.pdf}},
  \href {http://dx.doi.org/10.1101/2020.03.23.20041608}
  {\path{doi:10.1101/2020.03.23.20041608}}.
\newline\urlprefix\url{https://www.medrxiv.org/content/early/2020/03/27/2020.03.23.20041608}

\bibitem{barabasi2016network}
A.~L. Barabasi, M.~Posfai,
  \href{http://barabasi.com/networksciencebook/}{Network science}, Cambridge
  University Press, 2016.
\newline\urlprefix\url{http://barabasi.com/networksciencebook/}

\bibitem{LI2020282}
L.~Li, Z.~Yang, Z.~Dang, C.~Meng, J.~Huang, H.~Meng, D.~Wang, G.~Chen,
  J.~Zhang, H.~Peng, Y.~Shao,
  \href{http://www.sciencedirect.com/science/article/pii/S2468042720300087}{Propagation
  analysis and prediction of the covid-19}, Infectious Disease Modelling 5
  (2020) 282 -- 292.
\newblock \href {http://dx.doi.org/https://doi.org/10.1016/j.idm.2020.03.002}
  {\path{doi:https://doi.org/10.1016/j.idm.2020.03.002}}.
\newline\urlprefix\url{http://www.sciencedirect.com/science/article/pii/S2468042720300087}

\bibitem{ZHUANG2020308}
Z.~Zhuang, S.~Zhao, Q.~Lin, P.~Cao, Y.~Lou, L.~Yang, S.~Yang, D.~He, L.~Xiao,
  \href{http://www.sciencedirect.com/science/article/pii/S1201971220302599}{Preliminary
  estimates of the reproduction number of the coronavirus disease (covid-19)
  outbreak in republic of korea and italy by 5 march 2020}, International
  Journal of Infectious Diseases 95 (2020) 308 -- 310.
\newblock \href {http://dx.doi.org/https://doi.org/10.1016/j.ijid.2020.04.044}
  {\path{doi:https://doi.org/10.1016/j.ijid.2020.04.044}}.
\newline\urlprefix\url{http://www.sciencedirect.com/science/article/pii/S1201971220302599}

\bibitem{Ziff2020.02.16.20023820}
A.~L. Ziff, R.~M. Ziff,
  \href{https://www.medrxiv.org/content/early/2020/03/03/2020.02.16.20023820}{Fractal
  kinetics of covid-19 pandemic}, medRxiv\href
  {http://arxiv.org/abs/https://www.medrxiv.org/content/early/2020/03/03/2020.02.16.20023820.full.pdf}
  {\path{arXiv:https://www.medrxiv.org/content/early/2020/03/03/2020.02.16.20023820.full.pdf}},
  \href {http://dx.doi.org/10.1101/2020.02.16.20023820}
  {\path{doi:10.1101/2020.02.16.20023820}}.
\newline\urlprefix\url{https://www.medrxiv.org/content/early/2020/03/03/2020.02.16.20023820}

\bibitem{li2020scaling}
M.~Li, J.~Chen, Y.~Deng, Scaling features in the spreading of covid-19 (2020).
\newblock \href {http://arxiv.org/abs/2002.09199} {\path{arXiv:2002.09199}}.

\bibitem{Herrmann2020.04.02.20050468}
H.~A. Herrmann, J.~Schwartz,
  \href{https://www.medrxiv.org/content/early/2020/04/06/2020.04.02.20050468}{Using
  network science to propose strategies for effectively dealing with pandemics:
  The covid-19 example}, medRxiv\href
  {http://arxiv.org/abs/https://www.medrxiv.org/content/early/2020/04/06/2020.04.02.20050468.full.pdf}
  {\path{arXiv:https://www.medrxiv.org/content/early/2020/04/06/2020.04.02.20050468.full.pdf}},
  \href {http://dx.doi.org/10.1101/2020.04.02.20050468}
  {\path{doi:10.1101/2020.04.02.20050468}}.
\newline\urlprefix\url{https://www.medrxiv.org/content/early/2020/04/06/2020.04.02.20050468}

\bibitem{PhysRevE.65.055103}
Z.~Dezs\ifmmode~\mbox{\H{o}}\else \H{o}\fi{}, A.-L. Barab\'asi,
  \href{https://link.aps.org/doi/10.1103/PhysRevE.65.055103}{Halting viruses in
  scale-free networks}, Phys. Rev. E 65 (2002) 055103.
\newblock \href {http://dx.doi.org/10.1103/PhysRevE.65.055103}
  {\path{doi:10.1103/PhysRevE.65.055103}}.
\newline\urlprefix\url{https://link.aps.org/doi/10.1103/PhysRevE.65.055103}

\bibitem{Lacasa4972}
L.~Lacasa, B.~Luque, F.~Ballesteros, J.~Luque, J.~C. Nu{\~n}o,
  \href{https://www.pnas.org/content/105/13/4972}{From time series to complex
  networks: The visibility graph}, Proceedings of the National Academy of
  Sciences 105~(13) (2008) 4972--4975.
\newblock \href
  {http://arxiv.org/abs/https://www.pnas.org/content/105/13/4972.full.pdf}
  {\path{arXiv:https://www.pnas.org/content/105/13/4972.full.pdf}}, \href
  {http://dx.doi.org/10.1073/pnas.0709247105}
  {\path{doi:10.1073/pnas.0709247105}}.
\newline\urlprefix\url{https://www.pnas.org/content/105/13/4972}

\bibitem{demertzis2020modeling}
K.~Demertzis, D.~Tsiotas, L.~Magafas, Modeling and forecasting the covid-19
  temporal spread in greece: an exploratory approach based on complex network
  defined splines (2020).
\newblock \href {http://arxiv.org/abs/2005.01163} {\path{arXiv:2005.01163}}.

\bibitem{Pujari2020.03.13.20035386}
B.~S. Pujari, S.~M. Shekatkar,
  \href{https://www.medrxiv.org/content/early/2020/03/17/2020.03.13.20035386}{Multi-city
  modeling of epidemics using spatial networks: Application to 2019-ncov
  (covid-19) coronavirus in india}, medRxiv\href
  {http://arxiv.org/abs/https://www.medrxiv.org/content/early/2020/03/17/2020.03.13.20035386.full.pdf}
  {\path{arXiv:https://www.medrxiv.org/content/early/2020/03/17/2020.03.13.20035386.full.pdf}},
  \href {http://dx.doi.org/10.1101/2020.03.13.20035386}
  {\path{doi:10.1101/2020.03.13.20035386}}.
\newline\urlprefix\url{https://www.medrxiv.org/content/early/2020/03/17/2020.03.13.20035386}

\bibitem{biswas2020spacetime}
K.~Biswas, P.~Sen, Space-time dependence of corona virus (covid-19) outbreak
  (2020).
\newblock \href {http://arxiv.org/abs/2003.03149} {\path{arXiv:2003.03149}}.

\bibitem{Gross2020}
B.~Gross, Z.~Zheng, S.~Liu, X.~Chen, A.~Sela, J.~Li, D.~Li, S.~Havlin,
  \href{https://www.medrxiv.org/content/early/2020/04/11/2020.03.23.20041517}{Spatio-temporal
  propagation of covid-19 pandemics}, medRxiv\href
  {http://arxiv.org/abs/https://www.medrxiv.org/content/early/2020/04/11/2020.03.23.20041517.full.pdf}
  {\path{arXiv:https://www.medrxiv.org/content/early/2020/04/11/2020.03.23.20041517.full.pdf}},
  \href {http://dx.doi.org/10.1101/2020.03.23.20041517}
  {\path{doi:10.1101/2020.03.23.20041517}}.
\newline\urlprefix\url{https://www.medrxiv.org/content/early/2020/04/11/2020.03.23.20041517}

\bibitem{PokkuluriDeviNedunuri2020}
K.~S. Pokkuluri, N.~D. Nedunuri, U.~S. Usha, A novel cellular automata
  classifier for covid-19 prediction, Journal of Health Sciences 10~(1) (2020)
  34--38.

\bibitem{Chinamigrant}
C.~Fan, T.~Cai, Z.~Gai, Y.~Wu, The relationship between the migrant population
  migration network and the risk of covid 19 transmission in china empirical
  analysis and prediction in prefecture level cities, International Journal of
  Environmental Research and Public Health 17.

\bibitem{naturecommunication}
F.~S. Lu, M.~W. Hattab, C.~L. Clemente, M.~Biggerstaff, M.~Santillana, Improved
  state-level influenza nowcasting in the united states leveraging
  internet-based data and network approaches, Nature Communications~(10).

\bibitem{Quinsocialijerph17072365}
L.~Qin, Q.~Sun, Y.~Wang, K.~Wu, M.~Chen, B.~Shia, S.~Wu,
  \href{https://www.mdpi.com/1660-4601/17/7/2365}{Prediction of number of cases
  of 2019 novel coronavirus (covid-19) using social media search index},
  International Journal of Environmental Research and Public Health 17~(7).
\newblock \href {http://dx.doi.org/10.3390/ijerph17072365}
  {\path{doi:10.3390/ijerph17072365}}.
\newline\urlprefix\url{https://www.mdpi.com/1660-4601/17/7/2365}

\bibitem{twitterIranauthor}
K.~Jahanbin, V.~Rahmanian, Using twitter and web news mining to predict
  covid-19 outbreak, Asian Pacific Journal of Tropical Medicine\href
  {http://arxiv.org/abs/http://www.apjtm.org/preprintarticle.asp?id=279651}
  {\path{arXiv:http://www.apjtm.org/preprintarticle.asp?id=279651}}.

\bibitem{liu2020machine}
D.~Liu, L.~Clemente, C.~Poirier, X.~Ding, M.~Chinazzi, J.~T. Davis,
  A.~Vespignani, M.~Santillana, A machine learning methodology for real-time
  forecasting of the 2019-2020 covid-19 outbreak using internet searches, news
  alerts, and estimates from mechanistic models (2020).
\newblock \href {http://arxiv.org/abs/2004.04019} {\path{arXiv:2004.04019}}.

\bibitem{info:doi/10.2196/18828}
S.~M. Ayyoubzadeh, S.~M. Ayyoubzadeh, H.~Zahedi, M.~Ahmadi,
  S.~R~Niakan~Kalhori, Predicting covid-19 incidence through analysis of google
  trends data in iran: Data mining and deep learning pilot study, JMIR Public
  Health Surveillance 6~(2) (2020) e18828.
\newblock \href {http://dx.doi.org/10.2196/18828} {\path{doi:10.2196/18828}}.

\bibitem{lampos2020tracking}
V.~Lampos, S.~Moura, E.~Yom-Tov, M.~Edelstein, M.~Majumder, Y.~Hamada, M.~X.
  Rangaka, R.~A. McKendry, I.~J. Cox, Tracking covid-19 using online search
  (2020).
\newblock \href {http://arxiv.org/abs/2003.08086} {\path{arXiv:2003.08086}}.

\bibitem{twitterdubey}
A.~D. Dubey, Twitter sentiment analysis during covid19 outbreak (2020).
\newblock \href {http://arxiv.org/abs/Available at SSRN:
  https://ssrn.com/abstract=3572023} {\path{arXiv:Available at SSRN:
  https://ssrn.com/abstract=3572023}}.

\bibitem{PETROSILLO2020729}
N.~Petrosillo, G.~Viceconte, O.~Ergonul, G.~Ippolito, E.~Petersen,
  \href{http://www.sciencedirect.com/science/article/pii/S1198743X20301713}{Covid-19,
  sars and mers: are they closely related?}, Clinical Microbiology and
  Infection 26~(6) (2020) 729 -- 734.
\newblock \href {http://dx.doi.org/https://doi.org/10.1016/j.cmi.2020.03.026}
  {\path{doi:https://doi.org/10.1016/j.cmi.2020.03.026}}.
\newline\urlprefix\url{http://www.sciencedirect.com/science/article/pii/S1198743X20301713}

\bibitem{covidtesting}
Coronavirus cases, \url{https://www.worldometers.info/coronavirus/#countries},
  accessed 31st May, 2020.

\bibitem{mild}
Coronavirus disease 2019 (covid-19) situation report 46,
  \url{https://www.who.int/docs/default-source/coronaviruse/situation-reports/20200306-sitrep-46-covid-19.pdf?sfvrsn=96b04adf_4},
  accessed 31st May, 2020.

\bibitem{Scialstigma}
Social stigma associated with covid-19,
  \url{https://www.who.int/docs/default-source/coronaviruse/covid19-stigma-guide.pdf?sfvrsn=226180f4_2},
  accessed 31st May, 2020.

\bibitem{Logisticno}
M.~A.~A. Bastian~Prasse, P.~V. Mieghem, Fundamental limits of predicting
  epidemic outbreaks,
  \url{https://www.nas.ewi.tudelft.nl/people/Piet/papers/TUD2020410_prediction_limits_epidemic_outbreaks.pdf},
  accessed on $20^{th}$ May 2020.

\bibitem{internetusers}
Cia world factbook,
  \url{https://www.cia.gov/library/publications/the-world-factbook/rankorder/2153rank.html},
  accessed 26th May, 2020.

\bibitem{DBLP:conf/www/ZouLC19}
B.~Zou, V.~Lampos, I.~J. Cox,
  \href{https://doi.org/10.1145/3308558.3313477}{Transfer learning for
  unsupervised influenza-like illness models from online search data}, in:
  L.~Liu, R.~W. White, A.~Mantrach, F.~Silvestri, J.~J. McAuley,
  R.~Baeza{-}Yates, L.~Zia (Eds.), The World Wide Web Conference, {WWW} 2019,
  San Francisco, CA, USA, May 13-17, 2019, {ACM}, 2019, pp. 2505--2516.
\newblock \href {http://dx.doi.org/10.1145/3308558.3313477}
  {\path{doi:10.1145/3308558.3313477}}.
\newline\urlprefix\url{https://doi.org/10.1145/3308558.3313477}

\end{thebibliography}

\end{document}